\documentclass[twocolumn,epjc3,final]{svjour3}
\RequirePackage[T1]{fontenc}

\smartqed  % flush right qed marks, e.g. at end of proof

\RequirePackage{graphicx}
\RequirePackage{mathptmx}      % use Times fonts if available on your TeX system
\RequirePackage{flushend}
\RequirePackage[numbers,sort&compress]{natbib}

\usepackage{amsmath, txfonts,multirow}
\usepackage{graphicx,bm,booktabs}
\usepackage{color}
\usepackage{multirow}
\usepackage[bookmarks=true,bookmarksopen=false,plainpages=false,breaklinks=true,
   bookmarksnumbered=true,hypertexnames=false,
   filecolor=blue,urlcolor=three,menucolor=three,
   linkcolor=three,citecolor=blueone, colorlinks,
   anchorcolor=blue,runcolor=pink,frenchlinks=red
   pdfstartview=FitH,pdftitle=title,%
   pdfauthor=author]{hyperref}
   
\allowdisplaybreaks[4]

\definecolor{cover}{rgb}{0.77,0.87,0.88}
\definecolor{blueone}{rgb}{0.1,0.1,.7}
\definecolor{citec}{rgb}{0.14,0.47,0.09}
\definecolor{two}{rgb}{0.0,0.5,0.}
\definecolor{three}{rgb}{.5,.1,0.15}

\begin{document}

\title{Possible  molecular dibaryons with $csssqq$ quarks and their baryon-antibaryon partners}
\author{Shu-Yi Kong\thanksref{addr1}, Jun-Tao Zhu\thanksref{addr2}, Jun He\thanksref{addr1,addr2,e1}}
\thankstext{e1}{Corresponding author: junhe@njnu.edu.cn}
\institute{School of Physics and Technology, Nanjing Normal University, Nanjing 210097, China\label{addr1}
\and
Lanzhou Center for Theoretical Physics, Lanzhou University, Lanzhou 730000, China\label{addr2}
}

\date{Received: date / Revised version: date}

\maketitle
\begin{abstract}

In this work, we systematically investigate the charmed-strange dibaryon systems
with $csssqq$ quarks and their baryon-antibaryon partners from the interactions
$\Xi^{(',*)}_{c}\Xi^{(*)}$, $\Omega^{(*)}_c\Lambda$,
$\Omega^{(*)}_c\Sigma^{(*)}$, $\Lambda_c\Omega$ and $\Sigma^{(*)}_c\Omega$ and
their baryon-antibaryon partners from interactions
$\Xi^{(',*)}_{c}\bar{\Xi}^{(*)}$, $\Omega^{(*)}_c\bar{\Lambda}$,
$\Omega^{(*)}_c\bar{\Sigma}^{(*)}$, $\Lambda_c\bar{\Omega}$ and
$\Sigma^{(*)}_c\bar{\Omega}$.  The potential kernels are constructed with the
help of effective Lagrangians under SU(3), heavy quark, and chiral symmetries to
describe these  interactions. To search for  possible molecular states, the
kernels are inserted into the quasipotential Bethe-Salpeter equation, which is
solved to find poles from scattering amplitude. The results suggest that 36 and
24 bound states can be found in the baryon-baryon and  baryon-antibaryon
interactions, respectively. However, much large values of parameter $\alpha$ are
required to produce the bound states from the baryon-antibaryon interactions,
which questions the existence of these bound states. Possible coupled-channel
effect are considered in the current work to estimate the couplings of the
molecular states to the channels considered.

\end{abstract}

\section{Introduction}

As an important type of exotic hadrons, the dibaryons  with baryon quantum
number $B=2$ attract much attention from the hadron physics community. In fact,
one type of the exotic hadrons proposed earliest in the literature is the
dibaryons predicted by Dyson and Xuong in 1964 based on the SU(6) symmetry
almost at the same time of the proposal of the quark model~\cite{Dyson:1964xwa}.
The WASA-at-COSY collaboration reported a new resonance $d^*(2380)$ with quantum
number $I(J^P)=0(3^+)$, a mass of about 2370 MeV, and a width of about 70 MeV in
the process $pp\rightarrow d\pi^0\pi^0$ at ~\cite{WASA-at-COSY:2011bjg}. Soon
after the observation of the  $d^*(2380)$, it is related to the dibaryon
predicted~\cite{Gal:2013dca,Huang:2014kja} while there is still other
interpretations, such as a triangle singularity in the last step of the reaction
in a sequential single pion production process~\cite{Ikeno:2021frl}. More
experimental and theoretical works are still required to clarify its origin.

These early proposed dibaryons are exotic hadrons in the light flavor sector. In
the past decades,  many candidates of exotic states in charmed sector, such as
hidden-charm tetraquarks and pentaquarks, have been observed in experiment, for
example, the $X(3872)$ and
$Z_c(3900)$~\cite{Belle:2003nnu,Tornqvist:2004qy,BESIII:2013ris,Belle:2013yex,Xiao:2013iha},
and a series of hidden-charm pentaquarks $P_c$~\cite{LHCb:2015yax,LHCb:2019kea,LHCb:2020jpq,LHCb:2022jad}. These states
were observed near the thresholds of two charmed hadrons. Hence, it is natural
to interpret them as the molecular states produced from interactions of a pair
of charm and anticharm hadrons. Motivated by the observations of these states,
theorists expect that there may exist dibaryon molecules composed of two heavy
baryons.  Due to large masses of the heavy baryons, the kinetic energy of a
dibaryon system is reduced, which makes it easier to form a bound state.
Possible hidden-charm and double-charm dibaryons were investigated in different
approaches~\cite{Carames:2015sya,Lu:2022myk,
Vijande:2016nzk,Meng:2017fwb,Li:2012bt,Dong:2021juy,Chen:2021cfl,Liu:2021gva,Song:2022svi}.
These results suggest that attraction may exist between a charmed baryon and an
anticharmed or charmed baryon by light meson exchanges, which favors the
existence of hidden-charm  dibaryon molecular states and their double-charm
partners.

In addition to the above hidden-charm and double-charm states, some
charmed-strange states were also observed these years, and taken as the candidates of
molecular states of a charmed meson and a strange meson  in the literature. As
early as 2003, the BaBar collaboration reported a narrow peak $D^{*}_{s0}(2317)$
near the $DK$ threshold~\cite{BaBar:2003oey}, and later confirmed at  CLEO and
BELLE~\cite{CLEO:2003ggt,Belle:2003guh}.  The CLEO collaboration also observed
another narrow peak, the $D_{s1}(2460)$  near the $D^*K$
threshold~\cite{CLEO:2003ggt}. These states can not be well put into the
conventional quark model with a charmed and an antistrange quark. Since these
charmed-strange states are very close to the threshold of a charmed meson and a
strange meson, some authors interpreted them as the molecules of
corresponding charmed and strange
mesons~\cite{Barnes:2003dj,Oset:2016cnj,Kolomeitsev:2003ac,Guo:2006fu,Guo:2006rp,Rosner:2006vc,
Zhang:2006ix,Liu:2020nil}. Recently the LHCb collaboration reported the $X_0(2900)$ and $X_1(2900)$ near the $\bar{D}^{*}K^{*}$
threshold~\cite{LHCb:2020bls,LHCb:2020pxc}. Such states should be composed of
four different quarks, and soon be explained as $\bar{D}^*K^*$ molecular
state~\cite{Agaev:2020nrc,Huang:2020ptc,Mutuk:2020igv,Molina:2020hde,Xiao:2020ltm,He:2020btl,Kong:2021ohg}.
By adding an additional light quark to the above charmed-strange molecular
states, the existence of charmed-strange pentaquark molecular states were also
predicted in Refs.~\cite{Liu:2022qns,An:2022vtg,Chen:2022svh}.

Following this way, if we continue to add light quark and convert all antiquarks
to quarks,  we will reach a charm-strange dibaryon systems.  In
Ref.~\cite{Kong:2022rvd}, we systematically investigated the charmed-strange
dibaryons with $csqqqq$ quarks and their baryon-antibaryon partners from the
interactions of a charmed baryon and a strange baryon $\Lambda_c\Lambda$,
$\Lambda_c\Sigma^{(*)}$, $\Sigma_c^{(*)}\Lambda$, and
$\Sigma^{(*)}_c\Sigma^{(*)}$, and corresponding interactions of a charmed baryon
and an antistrange baryon $\Lambda_c\bar{\Lambda}$,
$\Lambda_c\bar{\Sigma}^{(*)}$, $\Sigma^{(*)}_c\bar{\Lambda}$, and
$\Sigma^{(*)}_c\bar{\Sigma}^{(*)}$.  The calculation suggests that attractions
widely exist in charmed-strange dibaryon systems while few bound states are
produced from the charmed-antistrange interactions. If  one $u/d$ quark in
each constituent baryon is simultaneously replaced  by a strange quark,
we can reach charmed-strange dibaryon systems $\Xi^{(',*)}_{c}\Xi^{(*)}$, which
are scarcely studied in the literature. In this work, we will study these
systems together with the systems $\Omega^{(*)}_c\Lambda$,
$\Omega^{(*)}_c\Sigma^{(*)}$, $\Lambda_c\Omega$ and $\Sigma^{(*)}_c\Omega$ with
the same quark components, $csssqq$ quarks,  and their baryon-antibaryon
partners $\Xi^{(',*)}_{c}\bar{\Xi}^{(*)}$, $\Omega^{(*)}_c\bar{\Lambda}$,
$\Omega^{(*)}_c\bar{\Sigma}^{(*)}$, $\Lambda_c\bar{\Omega}$ and
$\Sigma^{(*)}_c\bar{\Omega}$.

The work is organized as follows. After introduction, the potential kernels of
systems considered are presented, which are obtained with the help of the
effective Lagrangians with SU(3), heavy quark, and chiral symmetries. The
quasipotential Bethe-Salpeter equation (qBSE) approach will also be introduced
briefly.  In Section \ref{Sec:BcB}, The bound states from all interactions will
be searched with single-channel calculations. In Section \ref{Sec:CCR}, the
bound states of the molecular states from full coupled-channel calculation will
be presented. And the poles from two-channel calculations are also provided to
estimate the strengths of the couplings between a molecular state and
corresponding channels. In Section \ref{Sec: summary}, discussion and summary
are given.

\section{Theoretical frame}\label{Sec: Formalism}

In this work, we consider the possible  molecular dibaryons from the
interactions $\Xi^{(',*)}_{c}\Xi^{(*)}$, $\Omega^{(*)}_c\Lambda$,
$\Omega^{(*)}_c\Sigma^{(*)}$, $\Lambda_c\Omega$ and $\Sigma^{(*)}_c\Omega$ and
their baryon-antibaryon partners $\Xi^{(',*)}_{c}\bar{\Xi}^{(*)}$,
$\Omega^{(*)}_c\bar{\Lambda}$, $\Omega^{(*)}_c\bar{\Sigma}^{(*)}$,
$\Lambda_c\bar{\Omega}$ and $\Sigma^{(*)}_c\bar{\Omega}$. The coupling between
different channels will also be included to make a coupled-channel calculation
to obtain the scattering amplitude by solving the qBSE.  To achieve this aim,
the potential will be constructed by the light meson exchanges. The Lagrangians
are required to obtain the vertices, and will be given below.

\subsection{ Relevant Lagrangians}

For the couplings of strange baryons with light mesons, we consider the exchange
of pseudoscalar mesons $P$ ($\pi$, $\eta$, $\rho$), vector mesons $V$ ($\omega$,
$\phi$, $K$, $K^*$), and $\sigma$ mesons. For the former seven mesons, the
vertices can be described by the effective Lagrangians with SU(3) and chiral
symmetries~\cite{Ronchen:2012eg,Kamano:2008gr}. The explicit the effective
Lagrangians reads,
\begin{eqnarray}
\mathcal{L}_{BBP}&=& -\frac{g_{BBP}}{m_P}\bar{B}\gamma^{5}\gamma^{\mu}\partial_{\mu}PB,\\
\mathcal{L}_{BBV}&=&-\bar{B}\left[g_{BBV}\gamma^{\mu}-\frac{f_{BBV}}{2m_{B}}\sigma^{\mu\nu}\partial_{\nu}\right]V_{\mu}B,\\
\mathcal{L}_{B^*B^*P}&=&\frac{g_{B^*B^*P}}{m_P}\bar{B^*}_{\mu}\gamma^{5}
\gamma^{\nu}B^{*\mu}\partial_{\nu}P, \\
\mathcal{L}_{B^*B^*V}&=&-\bar{B}^*_{\tau}\left[g_{B^*B^*V}\gamma^{\mu}-\frac{f_{B^*B^*V}}
{2m_{B^*}}\sigma^{\mu\nu}\partial_{\nu}\right]V_{\mu} B^{*\tau},\\
\mathcal{L}_{BB^*P}&=&\frac{g_{BB^*P}}{m_{P}}\bar{B}^{*\mu} \partial_{\mu}PB\;+\; \text{h.c.},\\
\mathcal{L}_{BB^*V}&=&-i \frac{g_{BB^*V}}{m_{V}}\bar{B}^{*\mu}\gamma^5\gamma^{\nu} V_{\mu\nu}B\;+\; \text{h.c.},
\end{eqnarray}
where $m_{p,V}$ is the mass of the pseudoscalar or vector meson. $B^{(*)}$ is the field of the strange baryon. $V_{\mu\nu}={\partial_{\mu}}\vec{V}_{\nu}-{\partial_{\mu}}\vec{V}_{\mu}$. The coupling constants can be determined by the SU(3) symmetry~\cite{Ronchen:2012eg,deSwart:1963pdg,Lu:2020qme,Zhu:2022fyb} with the coupling constants for the nucleon and $\Delta$. The SU(3) relations and the explicit values of coupling constants are calculated and listed in Table~\ref{constants}.

\renewcommand\tabcolsep{0.75cm}
\renewcommand{\arraystretch}{1.25}
\begin{table*}
\centering
\caption{The coupling constants in effective Lagrangians. Here,
$g_{BBP}=g_{NN\pi}=0.989$, $g_{BBV}=g_{NN\rho}=3.25$, $g_{B^*B^*P}=\sqrt{60} g_{\Delta\Delta\pi}=13.78$, $g_{B^*B^*V}=\sqrt{60} g_{\Delta\Delta\rho}=59.41$, $g_{BB^*P}=\sqrt{20} g_{N\Delta\pi}=9.48$, $g_{BB^*V}=\sqrt{20} g_{N\Delta\rho}=71.69$, $\alpha_{P}=0.4$, $\alpha_{V}=1.15$, $f_{NN\rho}= g_{NN\rho}\kappa_{\rho}$, $f_{\Delta\Delta\rho}= g_{\Delta\Delta\rho}\kappa_{\rho}$ with $\kappa_{\rho}=6.1$, $f_{NN\omega}=0$~\cite{Ronchen:2012eg,Lu:2020qme,Zhu:2022fyb}.\label{constants}}
\scalebox{0.99}{
\begin{tabular}[t]{lrr||lrr}\bottomrule[1.5pt]
 Coupling 				&SU(3) Relation		& Values			 				&Coupl.			&SU(3) Relation 		& Values	 \\ \hline
 $g_{\Xi\Xi\pi}$&$(2\alpha _P-1) g_{BBP}$&$-0.20$
& $g_{\Xi\Xi\eta}$&$-\frac{\sqrt{3}}{3}(1+2\alpha_P) g_{BBP}$&$-1.03$\\
$g_{\Xi\Xi\rho/\omega}$&$(2\alpha _V-1) g_{BBV}$&$4.23$
&$f_{\Xi\Xi\rho/\omega}$&$-\frac{1}{2}f_{NN\omega}-\frac{1}{2}f_{NN\rho}$&$-9.9$\\
$g_{\Xi\Xi\phi}$&$-2\sqrt{2}\alpha_V g_{BBV}$&$-10.57$
&$f_{\Xi\Xi\phi}$&$-\frac{\sqrt{2}}{2}f_{NN\omega}-\frac{\sqrt{2}}{2}f_{NN\rho}$&$-14.01$\\
$g_{\Xi^*\Xi^*\pi}$&$\frac{1}{4\sqrt{15}} g_{B^*B^*P}$&$0.89$
&$g_{\Xi^*\Xi^*\eta}$&$-\frac{1}{4\sqrt{5}} g_{B^*B^*P}$&$-1.54$\\
$g_{\Xi^*\Xi^*\rho/\omega}$&$\frac{1}{4\sqrt{15}} g_{B^*B^*V}$&$3.84$
&$f_{\Xi^*\Xi^*\rho/\omega}$&$\frac{1}{2}f_{\Delta\Delta\rho}$&$23.4$\\
$g_{\Xi^*\Xi^*\phi}$&$-\frac{1}{\sqrt{30}} g_{B^*B^*V}$&$-10.84$
&$f_{\Xi^*\Xi^*\phi}$&$-\sqrt{2}f_{\Delta\Delta\rho}$&$-66.16$\\
$g_{\Lambda\Lambda\omega}$ &$\frac{2}{3}(5\alpha_V-2)g_{BBV}$&$8.12$
 &$f_{\Lambda\Lambda\omega}$&$\frac{5}{6}f_{NN\omega}-\frac{1}{2}f_{NN\rho}$&$-9.9$\\
$g_{\Sigma\Sigma\pi}$&$2\alpha _P g_{BBP}$&$0.79$
& $g_{\Sigma\Sigma\eta}$&$\frac{2}{\sqrt{3}}(1-\alpha_P) g_{BBP}$&$0.68$\\
$g_{\Sigma\Sigma\rho/\omega}$&$2\alpha_V g_{BBV}$&$7.47$
&$f_{\Sigma\Sigma\rho/\omega}$&$\frac{1}{2}f_{NN\omega}+\frac{1}{2}f_{NN\rho}$&$9.9$\\
$g_{\Sigma^*\Sigma^*\pi}$&$\frac{1}{2\sqrt{15}} g_{B^*B^*P}$&$1.78$
&$g_{\Sigma^*\Sigma^*\eta}$&$0$&$0$\\
$g_{\Sigma^*\Sigma^*\rho/\omega}$&$\frac{1}{2\sqrt{15}} g_{B^*B^*V}$&$7.67$
&$f_{\Sigma^*\Sigma^*\rho/\omega}$&$f_{\Delta\Delta\rho}$&$46.78$\\
$g_{\Xi\Lambda K}$&$\frac{\sqrt{3}}{3}(4\alpha_P-1) g_{BBP}$&$0.34$
& $g_{\Xi\Sigma K}$&$-g_{BBP}$&$-0.98$\\
$g_{\Xi\Lambda K^*}$&$\frac{\sqrt{3}}{3}(4\alpha_P-1) g_{B^*B^*V}$&$6.75$
&$f_{\Xi\Lambda K^*}$&$\frac{\sqrt{3}}{3}f_{NN\omega}$&$0$\\
$g_{\Xi\Sigma K^*}$&$- g_{B^*B^*V}$&$-3.25$
&$f_{\Xi\Sigma K^*}$&$-f_{NN\rho}$&$-19.82$\\
$g_{\Xi^*\Sigma^*K}$&$-\frac{1}{2\sqrt{15}} g_{B^*B^*P}$&$-1.78$
&$g_{\Xi^*\Omega K}$&$\frac{1}{2\sqrt{10}} g_{B^*B^*P}$&$2.17$\\
$g_{\Xi^*\Sigma^*K^*}$&$-\frac{1}{2\sqrt{15}} g_{B^*B^*V}$&$-7.67$
&$f_{\Xi^*\Sigma^*K^*}$&-$f_{\Delta\Delta\rho}$&$46.78$\\
$g_{\Xi^*\Omega K^*}$&$\frac{1}{2\sqrt{10}} g_{B^*B^*V}$&$9.39$
&$f_{\Xi^*\Omega K^*}$&$\frac{\sqrt{6}}{2} f_{\Delta\Delta\rho}$&$57.29$\\
$g_{\Xi\Xi^*\pi}$&$\frac{1}{2\sqrt{30}}g_{BB^*P}$&$0.86$
&$g_{\Xi\Xi^*\eta}$&$-\frac{1}{2\sqrt{10}}g_{BB^*P}$&$-1.50$\\
$g_{\Xi\Xi^*\rho}$&$\frac{1}{2\sqrt{30}}g_{BB^*V}$&$6.54$
&$g_{\Xi\Xi^*\omega}$&$-\frac{1}{2\sqrt{30}}g_{BB^*V}$&$-6.54$\\
$g_{\Xi\Xi^*\phi}$&$-\frac{1}{2\sqrt{15}}g_{BB^*V}$&$-9.25$
&$g_{\Sigma\Sigma^*\pi}$&$\frac{1}{2\sqrt{30}}g_{BB^*P}$&$0.86$\\
$g_{\Sigma\Sigma^*\eta}$&$-\frac{1}{2\sqrt{10}}g_{BB^*P}$&$-1.49$
&$g_{\Sigma\Sigma^*\rho}$&$\frac{1}{2\sqrt{30}}g_{BB^*V}$&$6.54$\\
$g_{\Sigma\Sigma^*\omega}$&$-\frac{1}{2\sqrt{30}}g_{BB^*V}$&$-6.54$
&$g_{\Sigma\Sigma^*\phi}$&$-\frac{1}{2\sqrt{15}}g_{BB^*V}$&$-9.25$\\
$g_{\Xi^*\Lambda K}$&$\frac{1}{2\sqrt{10}}g_{BB^*P}$&$1.50$
&$g_{\Xi^*\Lambda K^*}$&$\frac{1}{2\sqrt{10}}g_{BB^*V}$&$11.34$\\
$g_{\Xi\Omega K}$&$\frac{1}{2\sqrt{5}}g_{BB^*P}$&$2.12$
&$g_{\Xi\Omega K^*}$&$\frac{1}{2\sqrt{5}}g_{BB^*V}$&$16.03$\\
$g_{\Xi^*\Sigma K}$&$-\frac{1}{2\sqrt{30}}g_{BB^*P}$&$-0.86$
&$g_{\Xi^*\Sigma K^*}$&$-\frac{1}{2\sqrt{30}}g_{BB^*V}$&$-6.54$\\
$g_{\Xi\Sigma^* K}$&$-\frac{1}{2\sqrt{30}}g_{BB^*P}$&$-0.86$
&$g_{\Xi\Sigma^* K^*}$&$-\frac{1}{2\sqrt{30}}g_{BB^*V}$&$-6.54$\\\bottomrule[1.5pt]
\end{tabular}
}
\end{table*}

For the couplings of strange baryons with the scalar meson $\sigma$, the
Lagrangians read~\cite{Zhao:2013ffn} \begin{eqnarray}
\mathcal{L}_{BB\sigma}&=&-g_{BB\sigma}\bar{B}\sigma B,\\
\mathcal{L}_{B^*B^*\sigma}&=&g_{B^*B^*\sigma}\bar{B}^{*\mu}\sigma B^*_{\mu}.
\end{eqnarray} The different choices of the mass of $\sigma$ meson from 400 to
550 MeV affects the result a little, which can be smeared by a small variation
of the cutoff in the calculation. In this work, we adopt a $\sigma$ mass of 500~MeV. In
general, we choose the coupling constants $g_{BB\sigma}$ and $g_{B^*B^*\sigma}$
as the same value as $g_{BB\sigma}=g_{B^*B^*\sigma}=6.59$~\cite{Zhao:2013ffn}.

For the couplings of charmed baryons with light mesons, the Lagrangians can be
constructed under the heavy quark and chiral
symmetries~\cite{Cheng:1992xi,Yan:1992gz,Wise:1992hn,Casalbuoni:1996pg}. The
explicit forms of the Lagrangians can be written as,
\begin{eqnarray}
{\cal L}_{BB\mathbb{P}}&=&-\frac{3g_1}{4f_\pi\sqrt{m_{\bar{B}}m_{B}}}~\epsilon^{\mu\nu\lambda\kappa}\partial^\nu \mathbb{P}~
\sum_{i=0,1}\bar{B}_{i\mu} \overleftrightarrow{\partial}_\kappa B_{j\lambda},\nonumber\\
{\cal L}_{BB\mathbb{V}}&=&-i\frac{\beta_S g_V}{2\sqrt{2m_{\bar{B}}m_{B}}}\mathbb{V}^\nu
 \sum_{i=0,1}\bar{B}_i^\mu \overleftrightarrow{\partial}_\nu B_{j\mu}\nonumber\\
&-&i\frac{\lambda_S
g_V}{\sqrt{2}}(\partial_\mu \mathbb{V}_\nu-\partial_\nu \mathbb{V}_\mu) \sum_{i=0,1}\bar{B}_i^\mu B_j^\nu,\nonumber\\
{\cal L}_{BB\sigma}&=&\ell_S\sigma\sum_{i=0,1}\bar{B}_i^\mu B_{j\mu},\nonumber\\
{\cal L}_{B_{\bar{3}}B_{\bar{3}}\mathbb{V}}&=&-i\frac{g_V\beta_B}{2\sqrt{2m_{\bar{B}_{\bar{3}}}m_{B_{\bar{3}}}} }\mathbb{V}^\mu\bar{B}_{\bar{3}}\overleftrightarrow{\partial}_\mu B_{\bar{3}},\nonumber\\
{\cal L}_{B_{\bar{3}}B_{\bar{3}}\sigma}&=&
\ell_B \sigma \bar{B}_{\bar{3}}B_{\bar{3}},\nonumber\\
{\cal L}_{BB_{\bar{3}}\mathbb{P}}
    &=&-i\frac{g_4}{f_\pi} \sum_i\bar{B}_i^\mu \partial_\mu \mathbb{P} B_{\bar{3}}+{\rm H.c.},\nonumber\\
{\cal L}_{BB_{\bar{3}}\mathbb{V}}    &=&\frac{g_\mathbb{V}\lambda_I}{\sqrt{2 m_{\bar{B}}m_{B_{\bar{3}}}}}\epsilon^{\mu\nu\lambda\kappa} \partial_\lambda \mathbb{V}_\kappa\sum_i\bar{B}_{i\nu} \overleftrightarrow{\partial}_\mu
   B_{\bar{3}}+{\rm H.c.},\label{LB}
\end{eqnarray}
where $m_{\bar{B},B,\bar{B}_3,B_3}$ is the mass of the charmed baryon. $S^{\mu}_{ab}$ is composed of the Dirac spinor operators,
\begin{eqnarray}
    S^{ab}_{\mu}&=&-\sqrt{\frac{1}{3}}(\gamma_{\mu}+v_{\mu})
    \gamma^{5}B^{ab}+B^{*ab}_{\mu}\equiv{ B}^{ab}_{0\mu}+B^{ab}_{1\mu},\nonumber\\
    \bar{S}^{ab}_{\mu}&=&\sqrt{\frac{1}{3}}\bar{B}^{ab}
    \gamma^{5}(\gamma_{\mu}+v_{\mu})+\bar{B}^{*ab}_{\mu}\equiv{\bar{B}}^{ab}_{0\mu}+\bar{B}^{ab}_{1\mu},
\end{eqnarray}
and the charmed baryon matrices are defined as,
\begin{align}
B_{\bar{3}}&=\left(\begin{array}{ccc}
0&\Lambda^+_c&\Xi_c^+\\
-\Lambda_c^+&0&\Xi_c^0\\
-\Xi^+_c&-\Xi_c^0&0
\end{array}\right),\quad
B=\left(\begin{array}{ccc}
\Sigma_c^{++}&\frac{1}{\sqrt{2}}\Sigma^+_c&\frac{1}{\sqrt{2}}\Xi'^+_c\\
\frac{1}{\sqrt{2}}\Sigma^+_c&\Sigma_c^0&\frac{1}{\sqrt{2}}\Xi'^0_c\\
\frac{1}{\sqrt{2}}\Xi'^+_c&\frac{1}{\sqrt{2}}\Xi'^0_c&\Omega^0_c
\end{array}\right), \nonumber\\
B^*&=\left(\begin{array}{ccc}
\Sigma_c^{*++}&\frac{1}{\sqrt{2}}\Sigma^{*+}_c&\frac{1}{\sqrt{2}}\Xi^{*+}_c\\
\frac{1}{\sqrt{2}}\Sigma^{*+}_c&\Sigma_c^{*0}&\frac{1}{\sqrt{2}}\Xi^{*0}_c\\
\frac{1}{\sqrt{2}}\Xi^{*+}_c&\frac{1}{\sqrt{2}}\Xi^{*0}_c&\Omega^{*0}_c
\end{array}\right).\label{MBB}
\end{align}
The $\mathbb P$ and $\mathbb V$ are the pseudoscalar and vector matrices as,
\begin{equation}
    {\mathbb P}=\left(\begin{array}{ccc}
        \frac{\sqrt{3}\pi^0+\eta}{\sqrt{6}}&\pi^+&K^+\\
        \pi^-&\frac{-\sqrt{3}\pi^0+\eta}{\sqrt{6}}&K^0\\
        K^-&\bar{K}^0&-\frac{2\eta}{\sqrt{6}}
\end{array}\right),
\mathbb{V}=\left(\begin{array}{ccc}
\frac{\rho^0+\omega}{\sqrt{2}}&\rho^{+}&K^{*+}\\
\rho^{-}&\frac{-\rho^{0}+\omega}{\sqrt{2}}&K^{*0}\\
K^{*-}&\bar{K}^{*0}&\phi
\end{array}\right).\nonumber \label{MPV}
\end{equation}

The parameters in the above Lagrangians are listed in Table~\ref{coupling}, which are cited from the literature~\cite{Chen:2019asm, Liu:2011xc, Isola:2003fh, Falk:1992cx}.
 \renewcommand\tabcolsep{0.425cm}
\renewcommand{\arraystretch}{1.2}
\begin{table}[h!]
\caption{The parameters and coupling constants. The $\lambda$, $\lambda_{S,I}$ and $f_\pi$ are in the unit of GeV$^{-1}$. Others are in the unit of $1$.
\label{coupling}}
\begin{tabular}{cccccccccccccccccc}\bottomrule[1.5pt]
%\\\hline
$f_\pi$&$g_V$&$\beta_S$&$\ell_S$&$g_1$\\
0.132 &5.9&-1.74&6.2&-0.94\\\hline
$\lambda_S$ &$\beta_B$&$\ell_B$ &$g_4$&$\lambda_I$\\
-3.31&$-\beta_S/2$&$-\ell_S/2$&$g_1/{2\sqrt{2}\over 3}$&$-\lambda_S/\sqrt{8}$ \\
\bottomrule[1.5pt]
\end{tabular}
\end{table}

\subsection{Potential kernel of interactions}

With the above Lagrangians for the vertices, the potential kernel can be
constructed in the one-boson-exchange model with the help of the standard
Feynman rule as in Refs.~\cite{He:2019ify,He:2015mja}. The propagators of the
exchanged light mesons are defined as, \begin{eqnarray}%
P_\mathbb{P,\sigma}(q^2)&=&
\frac{i}{q^2-m_\mathbb{P,\sigma}^2}~f_i(q^2),\nonumber\\
P^{\mu\nu}_\mathbb{V}(q^2)&=&i\frac{-g^{\mu\nu}+q^\mu
q^\nu/m^2_{\mathbb{V}}}{q^2-m_\mathbb{V}^2}~f_i(q^2), \end{eqnarray} where the
form factor $f_i(q^2)$ is adopted to reflect the off-shell effect of
exchanged meson, which is in form of $e^{-(m_e^2-q^2)^2/\Lambda_e^4}$ with $m_e$
and $q$ being the mass and momentum of the exchanged mesons,
respectively.

In this work, we still do not give the explicit form of the potential due to the
large number of channels to be considered. Instead, we input the vertices
$\Gamma$ obtained from the Lagrangians and the above propagators $P$ into the code directly. The
 dibaryon systems potential can be constructed with the help of
the standard Feynman rule as~\cite{He:2019ify}, \begin{eqnarray}%
{\cal V}_{\mathbb{P},\sigma}=I_{\mathbb{P},\sigma}\Gamma_1\Gamma_2
P_{\mathbb{P},\sigma}(q^2),\quad {\cal
V}_{\mathbb{V}}=I_{\mathbb{V}}\Gamma_{1\mu}\Gamma_{2\nu}
P^{\mu\nu}_{\mathbb{V}}(q^2),
\end{eqnarray}
where $I_{\mathbb{P},\mathbb{V},\sigma}$ is the flavor factors of the certain
meson exchange, which are listed in Table~\ref{flavor factor}.  The interaction
of their baryon-antibaryon partners interactions will be rewritten to the
charmed-strange interactions by the well-known G-parity rule
$V=\sum_{i}{\zeta_{i}V_{i}}$~\cite{PHILLIPS:1967wls,Klempt:2002ap}. The $G$
parities of the exchanged mesons $i$ are left as a $\zeta_{i}$ factor.  Since
$\pi$, $\omega$ and $\phi$ mesons carry odd $G$ parity, the $\zeta_{\pi}$,
$\zeta_{\omega}$ and $\zeta_{\phi}$ should equal $-1$, and others equal $1$.

\renewcommand\tabcolsep{0.09cm}
\renewcommand{\arraystretch}{1.42}
\begin{table}[h!]
 \centering
 \caption{The flavor factors $I_e$ for charmed-strange interactions.  The values for charmed-antistrange interactions can be obtained by G-parity rule from these of charmed-strange interactions. The  $I_\sigma$ should be 0 for coupling between different channels.\label{flavor factor}}
\begin{tabular}{lcccccccccc}\toprule[1.5pt]
&$I$& $\pi$& $\eta$ &$\rho$&$\omega$&$\phi$  &$\sigma$&$K$&$K^*$ \\\hline
$\Xi_c\Xi^{(*)}$-$\Xi_c\Xi^{(*)}$ &$0$&$-$&$-$&$-\frac{3\sqrt{2}}{2}$ &$\frac{\sqrt{2}}{2}$ & $1$&$2$&$-$&$-$\\
&$1$&$-$&$-$&$\frac{\sqrt{2}}{2}$ &$\frac{\sqrt{2}}{2}$ & $1$&$2$&$-$&$-$\\
$\Xi^{',*}_c\Xi^{(*)}$-$\Xi^{',*}_c\Xi^{(*)}$&$0$ &$-\frac{3\sqrt{2}}{4}$&$-\frac{1}{2\sqrt6}$&$-\frac{3\sqrt{2}}{4}$&$\frac{1}{2\sqrt{2}}$ &$\frac{1}{2}$ & $1$&$-$&$-$\\
&$1$&$\frac{\sqrt{2}}{4}$&$-\frac{1}{2\sqrt6}$&$\frac{\sqrt{2}}{4}$&$\frac{1}{2\sqrt{2}}$ &$\frac{1}{2}$ & $1$&$-$&$-$\\
$\Xi^{',*}_c\Xi^{(*)}$-$\Xi_c\Xi^{(*)}$&$0$ &$-\frac{3}{2}$&$\frac{\sqrt{3}}{2}$&$-\frac{3}{2}$&$\frac{1}{2}$ &$-\frac{1}{\sqrt{2}}$ & $-$&$-$&$-$\\
&$1$&$\frac{1}{2}$&$\frac{\sqrt{3}}{2}$&$\frac{1}{2}$&$\frac{1}{2}$ &$-\frac{1}{\sqrt{2}}$ & $-$&$-$&$-$\\
$\Omega^{(*)}_c\Lambda$-$\Omega^{(*)}_c\Lambda$&$0$&$-$ &$-\frac{2}{\sqrt{6}}$&$-$&$-$&$1$&$-$&$-$&$-$\\
$\Omega^{(*)}_c\Sigma^{(*)}$-$\Omega^{(*)}_c\Sigma^{(*)}$&$1$&$-$&$-\frac{2}{\sqrt{6}}$&$-$&$-$&$1$& $-$&$-$&$-$\\
$\Xi^{',*}_c\Xi^{(*)}$-$\Lambda_c\Omega$&$0$ &$-$&$-$&$-$&$-$ &$-$ & $-$&$-1$&$-1$\\
$\Xi_c\Xi^{(*)}$-$\Lambda_c\Omega$&$0$ &$-$&$-$&$-$&$-$ &$-$ & $-$&$-$&$\sqrt{2}$\\
$\Xi^{',*}_c\Xi^{(*)}$-$\Omega^{(*)}_c\Lambda$&$0$ &$-$&$-$&$-$&$-$ &$-$ & $-$&$-1$&$-1$\\
$\Xi_c\Xi^{(*)}$-$\Omega^{(*)}_c\Lambda$&$0$ &$-$&$-$&$-$&$-$ &$-$ & $-$&$-\sqrt{2}$&$-\sqrt{2}$\\
$\Xi^{',*}_c\Xi^{(*)}$-$\Sigma_c\Omega$&$1$ &$-$&$-$&$-$&$-$ &$-$ & $-$&$\frac{1}{\sqrt{2}}$&$\frac{1}{\sqrt{2}}$\\
$\Xi_c\Xi^{(*)}$-$\Sigma_c\Omega$&$1$ &$-$&$-$&$-$&$-$ &$-$ & $-$&$-1$&$-1$\\
$\Xi^{',*}_c\Xi^{(*)}$-$\Omega^{(*)}_c\Sigma^{(*)}$&$1$ &$-$&$-$&$-$&$-$ &$-$ & $-$&$-1$&$-1$\\
$\Xi_c\Xi^{(*)}$-$\Omega^{(*)}_c\Sigma^{(*)}$&$1$ &$-$&$-$&$-$&$-$ &$-$ & $-$&$-\sqrt{2}$&$-\sqrt{2}$\\
\bottomrule[1.5pt]
\end{tabular}
\end{table}

\subsection{The qBSE approach}

The Bethe-Salpeter equation is a 4-dimensional relativistic integral equation,
which can be used to treat two body scattering. In order to reduce the
4-dimensional Bethe-Salpeter equation to a 3-dimensional integral equation, we
adopt the covariant spectator approximation, which  keeps the unitary and
covariance of the equation~\cite{Gross:1991pm}. In such treatment, one of the
constituent particles, usually the heavier one, is put on shell, which leads to
a reduced propagator for two constituent particles in the center-of-mass frame
as~\cite{He:2015mja,He:2011ed},

\begin{eqnarray}
	G_0&=&\frac{\delta^+(p''^{~2}_h-m_h^{2})}{p''^{~2}_l-m_l^{2}}\nonumber\\
          &=&\frac{\delta^+(p''^{0}_h-E_h({\rm p}''))}{2E_h({\rm p''})[(W-E_h({\rm
p}''))^2-E_l^{2}({\rm p}'')]}.
\end{eqnarray}
As required by the spectator approximation adopted in the curren work, the heavier particle  ($h$ represents the charmed baryons) satisfies $p''^0_h=E_{h}({\rm p}'')=(m_{h}^{~2}+\rm p''^2)^{1/2}$. The $p''^0_l$ for the lighter particle (remarked as $l$) is then $W-E_{h}({\rm p}'')$. Here and hereafter, the value of the momentum  in center-of-mass frame is defined as ${\rm p}=|{\bm p}|$.

Then the 3-dimensional Bethe-Saltpeter
equation can be reduced to a 1-dimensional integral equation with fixed spin-parity $J^P$
by partial wave decomposition~\cite{He:2015mja},
\begin{eqnarray}
i{\cal M}^{J^P}_{\lambda'\lambda}({\rm p}',{\rm p})
&=&i{\cal V}^{J^P}_{\lambda',\lambda}({\rm p}',{\rm
p})+\sum_{\lambda''}\int\frac{{\rm
p}''^2d{\rm p}''}{(2\pi)^3}\nonumber\\
&\cdot&
i{\cal V}^{J^P}_{\lambda'\lambda''}({\rm p}',{\rm p}'')
G_0({\rm p}'')i{\cal M}^{J^P}_{\lambda''\lambda}({\rm p}'',{\rm
p}),\quad\quad \label{Eq: BS_PWA}
\end{eqnarray}
where the sum extends only over nonnegative helicity $\lambda''$.
The partial wave potential in 1-dimensional equation is defined with the potential
of the interaction obtained in the above as
\begin{eqnarray}
{\cal V}_{\lambda'\lambda}^{J^P}({\rm p}',{\rm p})
&=&2\pi\int d\cos\theta
~[d^{J}_{\lambda\lambda'}(\theta)
{\cal V}_{\lambda'\lambda}({\bm p}',{\bm p})\nonumber\\
&+&\eta d^{J}_{-\lambda\lambda'}(\theta)
{\cal V}_{\lambda'-\lambda}({\bm p}',{\bm p})],
\end{eqnarray}
where $\eta=PP_1P_2(-1)^{J-J_1-J_2}$ with $P$ and $J$ being parity and spin for
the system. The initial and final relative momenta are chosen as ${\bm
p}=(0,0,{\rm p})$ and ${\bm p}'=({\rm p}'\sin\theta,0,{\rm p}'\cos\theta)$. The
$d^J_{\lambda\lambda'}(\theta)$ is the Wigner d-matrix. Here, a regularization is
usually introduced to avoid divergence, when we treat an integral equation. In
the qBSE approach,  we usually adopt an exponential regularization by
introducing a form factor into the propagator as
$f(q^2)=e^{-(k_l^2-m_l^2)^2/\Lambda_r^4}$, where $k_l$ and $m_l$ are the
momentum and mass of the lighter one of and baryon.  In the current work, the
relation of the cutoff $\Lambda_r= m + \alpha_r$ 0.22~GeV with $m$ being the
mass of the exchanged meson is also introduced into the regularization form
factor as in those for the exchanged mesons. The cutoff  $\Lambda_e$ and $\Lambda_r$
play analogous roles in the calculation of the binding energy. For
simplification, we set  $\Lambda_e=\Lambda_r$ in the calculations.

The partial-wave qBSE is a one-dimensional integral equation, which can be
solved by discretizing the momenta with the Gauss quadrature.  It leads to a
matrix equation of a form $M=V+VGM$~\cite{He:2015mja}. The molecular state
corresponds to the pole of the amplitude, which can be obtained by varying $z$
to satisfy $|1-V(z)G(z)|=0$ where $z=E_R-i\Gamma/2$ being the exact position of
the bound state.

\section{Single-channel results}\label{Sec:BcB}

With previous information, the explicit numerical calculations will be performed
on the systems mentioned above. In the current model, we have the only one free
parameter $\alpha$. In the following, we vary the free parameter in a range of
0-5 to find the S-wave bound states with binding energy smaller than 30~MeV. In
this work, we consider all possible channels with $csssqq$ quarks, that is,
$\Xi^{(',*)}_{c}\Xi^{(*)}$, $\Omega^{(*)}_c\Lambda$,
$\Omega^{(*)}_c\Sigma^{(*)}$, $\Lambda_c\Omega$ and $\Sigma^{(*)}_c\Omega$ and
their baryon-antibaryon partners $\Xi^{(',*)}_{c}\bar{\Xi}^{(*)}$,
$\Omega^{(*)}_c\bar{\Lambda}$, $\Omega^{(*)}_c\bar{\Sigma}^{(*)}$,
$\Lambda_c\bar{\Omega}$ and $\Sigma^{(*)}_c\bar{\Omega}$. However, the
$\Lambda_c\Omega$, $\Sigma^{(*)}_c\Omega$ and their baryon-antibaryon partners
can not be considered in single-channel calculations due to the lack of 
exchanges of light mesons in the one-boson-exchange model considered in the
current work. However, these channels will  be considered  in the later
couple-channel calculations. Based on the quark configurations in different
hadron clusters, these single-channel interactions can be divided into two
categories: the $\Xi_c^{(',*)}\Xi^{(',*)}$ and $\Omega^{(*)}_c\Lambda$ or
$\Omega^{(*)}_c\Sigma^{(*)}$ and their baryon-antibaryon partners.

\subsection{Molecular states from interactions $\Xi_c^{(',*)}\Xi^{(',*)}$ and
$\Xi_c^{(',*)}\bar{\Xi}^{(',*)}$}

\begin{figure}[h!]
  \centering
  \includegraphics[scale=0.64,bb=34 27 408 500]{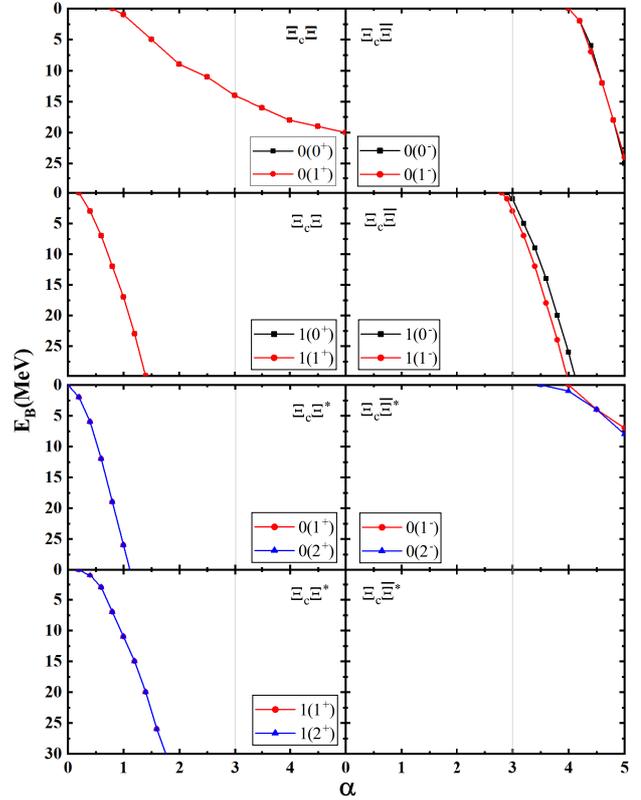}
  %\vspace{5em}
  \caption{Binding energies of bound states from the interactions $\Xi_c\Xi^{(*)}$ (left) and $\Xi_c\bar{\Xi}^{(*)}$(right) with thresholds of 3787 (4002) MeV  with the variation of $\alpha$ in single-channel calculation.}
  \label{b3b}
\end{figure}

First, we consider the interactions $\Xi_c^{(',*)}\Xi^{(*)}$ and
$\Xi_c^{(',*)}\bar{\Xi}^{(*)}$ with  quark configurations as $[csq][ssq]$ and
$[csq][\bar{s}\bar{s} \bar{q}]$, respectively.  The single-channel results for
the interactions $\Xi_c\Xi^{(*)}$ and $\Xi_c\bar{\Xi}^{(*)}$, in which the
charmed baryon belongs to the multiplet $B_{\bar{3}}$, are illustrated in
Fig~\ref{b3b}.  The results suggest that fourteen interactions produce bound
states in considered range of parameter $\alpha$. All eight bound states from
the $\Xi_c^{*}\Xi^{(*)}$ interaction can appear at $\alpha$ values less than 1. The
binding energies of the isovector $\Xi_c\Xi$ states with $(0,1)^+$ and the
isoscalar and isovector $\Xi_c\Xi^{*}$ states with $(1,2)^+$ both increase
rapidly to 30~MeV at $\alpha$ values of about 1.5, which indicates the strong
attraction.  However, the binding energies of isoscalar bound states  from the 
$\Xi_c\Xi$ interaction with $(1,2)^+$ increase slowly to 20~MeV at $\alpha$
values of about 5. The variation tendencies of the binding energies of the 
$\Xi_c\Xi^{(*)}$ states with different spin parities are analogous. Almost all
bound states from baryon-antibaryon interactions appear at the  $\alpha$ values 
more than 3 and the isovector $\Xi_c\bar{\Xi}^{*}$ interaction with $(1,2)^-$ can no produce bound state. It suggests that the possibility of the  existence of these
baryon-antibaryon bound states is relatively low.

\begin{figure}[h!] \centering
\includegraphics[scale=0.65,bb=34 27 408 500]{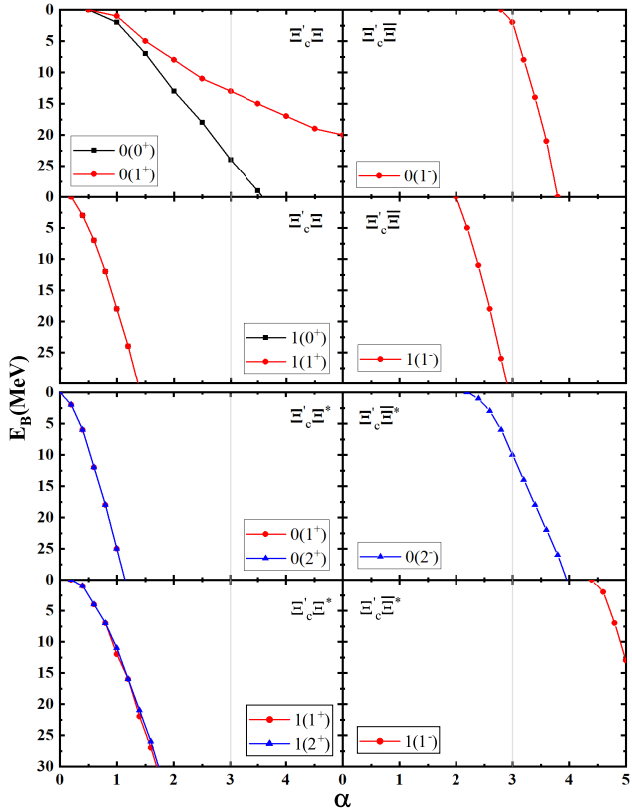}
%\vspace{5em}
\caption{Binding energies of  bound states from the interactions $\Xi^{'}_c\Xi^{(*)}$(left)
and $\Xi^{'}_c\bar{\Xi}^{(*)}$(right)  with thresholds of 3896 (4111) MeV with the variation of $\alpha$ in
single-channel calculation.} \label{b6} \end{figure}

In Fig.~\ref{b6}, the single-channel results about interactions
$\Xi^{'}_c\Xi^{(*)}$ and $\Xi^{'}_c\bar{\Xi}^{(*)}$ are presented.  In these
systems, the charmed baryon belongs to the multiplet $B_{6}$. The results
suggest that twelve bound states can be produced from these interactions within
considered range of  parameter $\alpha$. All eight bound states from
baryon-baryon interactions appear at $\alpha$ values less than 1. Among these
bound states, the two isoscalar bound states from the $\Xi^{'}_c\Xi$ interaction
with $(0,1)^+$ are well distinguished and  increase relatively slowly to 20~MeV
at $\alpha$ values of about 3.5 and 5.0, respectively. Other six bound states
increase rapidly to 30~MeV at $\alpha$ values of about 1.5, and the binding
energies for states with different spins are almost the same.  However, only
four bound states can be produced from baryon-antibaryon interactions, which
include the isoscalar and isovector $\Xi^{'}_c\bar{\Xi}$ states with $1^{-}$,
the isoscalar $\Xi^{'}_c\bar{\Xi}^{(*)}$ state with $2^{-}$, and isovector
$\Xi^{'}_c\bar{\Xi}^{(*)}$ state with $1^{-}$.  Again, one can still find that
the states with the larger spin are easy to be produced for the isoscalar
interactions, while the states with the smaller spin are easy to be produced for
the isovector interactions. Still, these baryon-antibaryon states are produced
at $\alpha$ values around or more than 3, which makes their coexistence less
possible.

In the following Fig.~\ref{b6a}, we present the results of the
$\Xi^{(*)}_c\Xi^{(*)}$ and $\Xi^{*}_c\bar{\Xi}^{(*)}$ systems, in which the
charmed baryon belongs to the multiplet $B_{6}^*$. The results suggest that
bound states can be produced from  eighteen interactions. For the baryon-baryon
systems, the bound states can be produced from all channels, and  appear at
$\alpha$ values below 1.5. The curves of two isoscalar $\Xi^{*}_c\Xi$ states
with $(0,1)^+$ are separated obviously, and their binding energies reach 5~MeV
relative slowly at $\alpha$ values about 4.5 and 2, respectively. Besides the two
states, other ten states increase with the parameter $\alpha$ to 30~MeV
relatively rapidly at $\alpha$ values of about 2.5.  Meanwhile, the interaction
with the smaller spins have stronger attractions, which is reflected by the
binding energies increasing faster with the variation of parameter. For their
baryon-anibaryon partners, two  isoscalar states from the
$\Xi^{*}_c\bar{\Xi}$ interaction  with $2^-$ and interaction
$\Xi^{*}_c\bar{\Xi}^*$ with $3^-$, as well as  four isovector states from the
$\Xi^{*}_c\bar{\Xi}$ interaction with $(1,2)^-$ and the $\Xi^{*}_c\bar{\Xi}^*$
interaction with $(0,1)^-$, can be produced at the cutoff over 2.5.

\begin{figure}[h!] \centering
  \includegraphics[scale=0.65,bb=34 27 408 500]{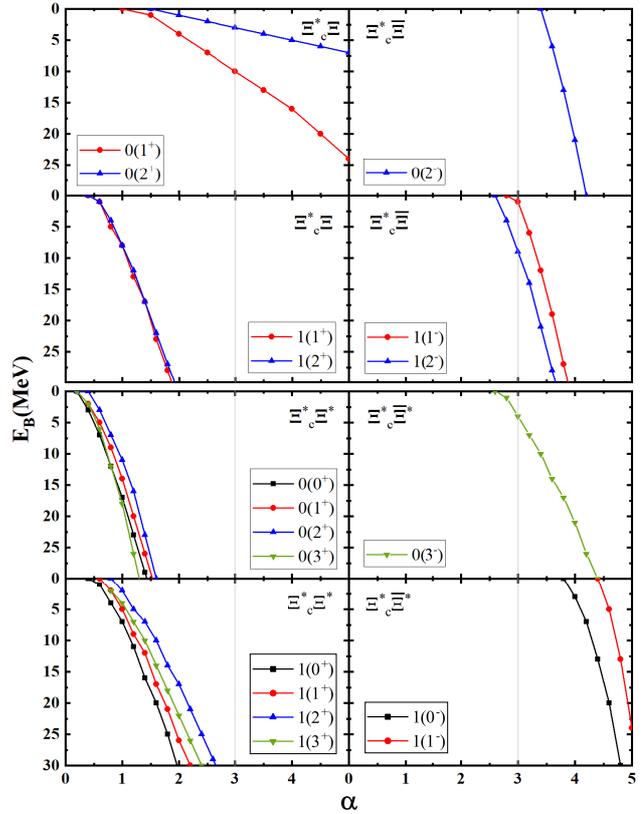}
  %\vspace{5em}
  \caption{Binding energies of  bound states from the interactions $\Xi^{*}_c\Xi^{(*)}$(left)
  and $\Xi^{*}_c\bar{\Xi}^{(*)}$(right)  with thresholds of 3963 (4178) MeV with the variation of $\alpha$ in
  single-channel calculation.} \label{b6a}
\end{figure}

\subsection{Molecular states from  interactions
$\Omega^{(*)}_c\Lambda/\Omega^{(*)}_c\Sigma^{(*)}$ and
$\Omega^{(*)}_c\bar{\Lambda}/\Omega^{(*)}_c\bar{\Sigma}^{(*)}$}

\begin{figure}[h!]
  \centering
  \includegraphics[scale=0.65,bb=402 145 34 508]{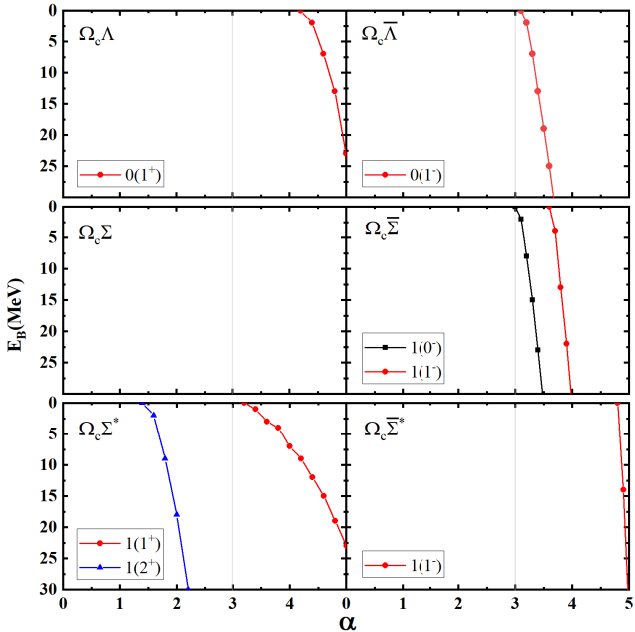}
  %\vspace{5em}
  \caption{Binding energies of  bound states from the interactions $\Omega_c\Lambda/\Omega_c\Sigma^{(*)}$(left) and $\Omega_c\bar{\Lambda}/\Omega_c\bar{\Sigma}^{(*)}$(right)  with thresholds of 3810/3888 (4079) MeV  with the variation of $\alpha$ in single-channel calculation.}
  \label{omegab6}
\end{figure}

For the systems composed of $[css][sqq]$ and $[css][\bar{s}\bar{q}\bar{q}]$,
there exist interactions $\Omega^{(*)}_c\Lambda$, $\Omega^{(*)}_c\Sigma^{(*)}$
and their baryon-antibaryon partners, interactions $\Omega^{(*)}_c\bar{\Lambda}$
and $\Omega^{(*)}_c\bar{\Sigma}^{(*)}$.  In Fig.~\ref{omegab6}, we first give
the results about  the interactions $\Omega_c\Lambda$, $\Omega_c\Sigma^{(*)}$,
$\Omega_c\bar{\Lambda}$ and $\Omega_c\bar{\Sigma}^{(*)}$, in which the charmed
baryons belong the multiplet $B_6$. Only seven states are produced from those
interactions. For the $\Omega_c\Lambda$ interaction and its baryon-antibaryon
partner $\Omega_c\bar{\Lambda}$ with isospin $I=0$, only the states that spin
$J=1$ can be produced at the cutoff about 4.0 and 3.0, respectively. There is no
bound state produced from the isovector interaction $\Omega_c\Sigma$ with
$(0,1)^+$ in the considered range of the parameter $\alpha$. Two bound states
from the $\Omega_c\bar{\Sigma}$ interaction with $(0,1)^-$ appear at $\alpha$
values of about 3.0 and 3.6, respectively. Two bound states from the isovector
$\Omega_c\Sigma^*$ interaction with $(1,2)^+$ appear at $\alpha$ value of about
3.0 and 1.5, respectively, while only an isovector $\Omega_c\bar{\Sigma}^*$
state with $1^-$ can be produced at  $\alpha$ value of about 4.8. The states
from the baryon-antibaryon interactions are still less likely to coexistence due
to the large values of parameter $\alpha$ required to produce the bound states.

\begin{figure}[h!]
  \centering
  \includegraphics[scale=0.65,bb=33 140 399 505]{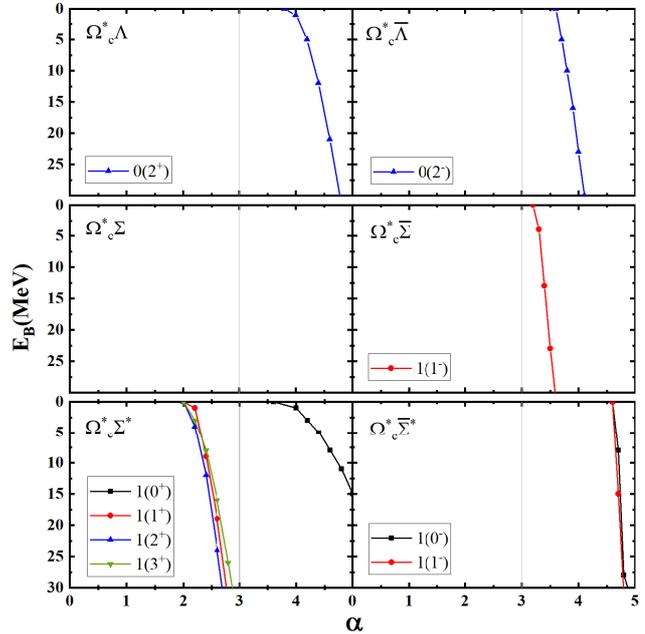}
  %\vspace{5em}
  \caption{Binding energies of  bound states from the $\Omega^*_c\Lambda/\Omega^*_c\Sigma^{(*)}$(left) and $\Omega^*_c\bar{\Lambda}/\Omega^*_c\bar{\Sigma}^{(*)}$(right)  with thresholds of 3882/3959 (4150) MeV with the variation of $\alpha$ in single-channel calculation.}
  \label{omegab6a}
\end{figure}

In Fig.~\ref{omegab6a}, the results about the interactions
$\Omega^{*}_c\Lambda$, $\Omega^{*}_c\Sigma^{(*)}$, $\Omega^{*}_c\bar{\Lambda}$, and $\Omega^{*}_c\bar{\Sigma}^{(*)}$ are presented.
Here, the charmed baryons are in the $B^*_6$ multiplet. The single-channel
calculation suggests that  nine bound states can be produced from sixteen
interactions considered. The isoscalar $\Omega^{*}_c\Lambda$ state and its
baryon-antibaryon partner $\Omega^{*}_c\bar{\Lambda}$ interaction with spin
$J=2$  appear at $\alpha$ of about 3.5. As the $\Omega_c\Sigma^{*}$ interaction,
the isovector $\Omega^{*}_c\Sigma$ systems with $(1,2)^+$ are unbound. The
$\Omega^{*}_c\bar{\Sigma}$ state with $1^-$ is produced at $\alpha$ larger than
3.0. The isovector interactions $\Omega^{*}_c\Sigma^{*}$ and
$\Omega^{*}_c\bar{\Sigma}^{*}$ are found attractive, and four states with spin
parities $(0,1,2,3)^+$ and two states with  $(0,1)^-$ are produced,
respectively. The $\Omega^{*}_ c\Sigma^{*}$ states with $0^+$ appear at $\alpha$
value of about 3.5, while the $(1,2,3)^+$ states all appear at cutoff about 2.0.
The two $\Omega^{*}_c\bar{\Sigma}^{*}$ with $(0,1)^-$ is produced at cutoff
about 4.6.

\section{Coupled-channel results}\label{Sec:CCR}

In the previous single-channel calculations, many bound states are produced from
the considered interactions  within allowed range of parameter $\alpha$.  To
estimate the strength of the coupling between a molecular state and the
corresponding decay channels, we will consider the couple-channel effects. In
the coupled-channel calculations, the channels with the same quark components
and the same quantum numbers can couple to each other, which will make the pole
of the bound state deviate from the real axis to the complex energy plane and
acquire an imaginary part. The imaginary part corresponds to the state of the
width as $\Gamma=2{\rm Im}z$.  Here, we present the coupled-channel results of
the position of bound state as $M_{th}-z$ instead of the origin position $z$ of
the pole, with the $M_{th}$ being the nearest threshold.  In the above
single-channel calculations, much larger $\alpha$ values  are required to
produce the bound states from the baryon-antibaryon interactions, which suggests
that the possibility of the existence of these states are very low. Hence, in
the following coupled-channel calculations, we only consider the baryon-baryon
interactions.  In the Table.~\ref{t0}, we present the coupled-channel results of
the isoscalar baryon-baryon interactions, which involve all possible  couplings
between the channels $\Xi^{(',*)}_c\Xi^{(*)}$, $\Lambda_c\Omega$ and
$\Omega^{(*)}_c\Lambda$. The poles of full coupled-channel interaction under the
corresponding threshold with different $\alpha$ are given in the second and
third columns.

\renewcommand\tabcolsep{0.3cm}
\renewcommand{\arraystretch}{1.5}
\begin{table*}[hpbt!]   %hpbt!
\footnotesize
\caption{The masses and widths of isoscalar baryon-baryon molecular states at different values of $\alpha$. The ``$CC$" means full coupled-channel calculation.  The values of the complex position means mass of corresponding threshold subtracted by the position of a pole, $M_{th}-z$,  in the unit of MeV. The two short line "$--$" means the coupling does not exist. The imaginary part shown as "$0.0$" means too small value under the current precision chosen.\label{t0}}
%\caption{Coupled channel results of the baryon-baryon states with $I=1$\label{t1a}}
\begin{tabular}{crrrrrrrrrrrrrrr}\bottomrule[1.5pt]
\specialrule{0em}{1pt}{1pt}
$I=0$&$\alpha_r$ &\multicolumn{1}{c}{$CC$}& \multicolumn{1}{c}{$\Xi^{'}_c\Xi^*$} & \multicolumn{1}{c}{$\Xi_c\Xi^*$} & \multicolumn{1}{c}{$\Xi^{*}_c\Xi$} &  \multicolumn{1}{c}{  $\Lambda_c\Omega$}& \multicolumn{1}{c}{$\Xi^{'}_c\Xi$}& \multicolumn{1}{c}{$\Omega^*_c\Lambda$}&
\multicolumn{1}{c}{$\Omega_c\Lambda$}& \multicolumn{1}{c}{$\Xi_c\Xi$} \\
\specialrule{0em}{1pt}{1pt}
\hline
\specialrule{0em}{1pt}{1pt}
$\Xi^*_c\Xi^*(0^+)$
&$0.3$ &$1+0.4i$   &$1+0.0i$     &$1+0.1i$      &$1+0.0i$  &$1+0.0i$  &$1+0.3i$   &$1+0.0i$   &$1+0.2i$       &$1+0.0i$ \\
4178 {\rm MeV}
&$0.5$ &$5+0.8i$   &$5+0.1i$    &$5+0.1i$        &$5+0.1i$  &$5+0.0i$   &$5+0.8i$ &$5+0.1i$   &$5+0.6i$     &$5+0.1i$   \\
&$0.7$ &$11+1.5i$   &$10+0.2i$    &$10+0.1i$         &$9+0.2i$  &$9+0.0i$  &$9+1.7i$  &$9+0.1i$ &$10+1.0i$      &$9+0.2i$  \\
\specialrule{0em}{1pt}{1pt}
$\Xi^*_c\Xi^*(1^+)$
&$0.3$ &$0+0.6i$   &$0+0.1i$    &$0+0.2i$      &$0+0.2i$  &$0+0.0i$  &$0+0.2i$  &$0+0.1i$ &$0+0.1i$ &$0+0.0i$ \\
4178 {\rm MeV}
&$0.5$ &$3+1.4i$   &$3+0.2i$    &$5+0.1i$      &$3+0.5i$  &$3+0.0i$  &$3+0.7i$  &$3+0.4i$  &$3+0.5i$&$3+0.0i$   \\
&$0.7$ &$7+2.7i$   &$7+0.3i$    &$10+0.1i$      &$7+1.2i$  &$7+0.0i$  &$7+1.8i$  &$7+0.8i$ &$7+1.2i$&$7+0.2i$  \\
\specialrule{0em}{1pt}{1pt}
$\Xi^*_c\Xi^*(2^+)$
&$0.5$ &$1+1.0i$   &$1+0.3i$     &$1+0.0i$      &$1+0.6i$  &$1+0.0i$  &$1+0.2i$ &$1+0.4i$  &$1+0.1i$ &$2+0.0i$ \\
4178 {\rm MeV}
&$0.7$ &$4+1.8i$   &$4+0.6i$    &$5+0.1i$       &$4+1.5i$  &$5+0.0i$  &$5+0.5i$ &$5+1.0i$ &$5+0.3i$&$5+0.0i$   \\
&$0.9$ &$7+2.8i$   &$7+0.1i$    &$9+0.2i$       &$8+3.1i$  &$9+0.0i$  &$8+1.1i$ &$9+2.1i$  &$9+0.0i$&$9+0.1i$  \\
\specialrule{0em}{1pt}{1pt}
$\Xi^*_c\Xi^*(3^+)$
&$0.3$ &$0+0.6i$   &$1+0.2i$     &$1+0.1i$      &$0+0.1i$  &$1+0.0i$  &$0+0.1i$  &$0+0.1i$ &$0+0.0i$ &$0+0.1i$ \\
4178 {\rm MeV}
&$0.5$ &$4+1.9i$   &$4+0.5i$    &$4+0.1i$       &$4+0.4i$  &$4+0.0i$  &$4+0.2i$  &$4+0.4i$  &$4+0.1i$&$4+0.2i$   \\
&$0.7$ &$8+4.6i$   &$10+1.1i$    &$9+0.3i$       &$8+1.0i$  &$9+0.0i$  &$9+0.6i$ &$9+0.9i$  &$9+0.2i$&$14+1.2i$  \\
\specialrule{0em}{1pt}{1pt}
$\Xi^{'}_c\Xi^*(1^+)$
&$0.2$ &$2+0.7i$   &$--$     &$2+0.0i$      &$2+0.2i$  &$2+0.0i$  &$2+0.3i$  &$2+0.2i$ &$2+0.3i$ &$2+0.0i$ \\
4111 {\rm MeV}
&$0.4$ &$7+1.6i$   &$--$    &$6+0.0i$        &$6+0.5i$  &$7+0.0i$  &$6+0.8i$  &$6+0.4i$ &$6+0.6i$&$6+0.0i$   \\
&$0.6$ &$14+2.6i$   &$--$    &$12+0.1i$        &$12+1.1i$  &$12+0.0i$  &$12+1.6i$  &$12+0.8i$ &$12+1.1i$&$11+0.2i$  \\
\specialrule{0em}{1pt}{1pt}
$\Xi^{'}_c\Xi^*(2^+)$
&$0.2$ &$2+0.6i$   &$--$     &$2+0.0i$      &$2+0.2i$  &$2+0.0i$  &$2+0.2i$  &$2+0.2i$ &$2+0.1i$ &$2+0.0i$ \\
4111 {\rm MeV}
&$0.4$ &$7+1.7i$   &$--$    &$7+0.1i$       &$7+0.5i$  &$7+0.0i$   &$7+0.5i$ &$7+0.4i$   &$7+0.4i$&$7+0.0i$   \\
&$0.6$ &$14+3.7i$   &$--$    &$13+0.3i$       &$14+1.1$  &$13+0.0i$  &$13+1.2i$  &$14+0.8i$ &$13+1.0i$&$13+0.0i$  \\
\specialrule{0em}{1pt}{1pt}
$\Xi_c\Xi^*(1^+)$
&$0.2$ &$2+0.3i$   &$--$     &$--$      &$2+0.0i$  &$2+0.0i$  &$2+0.1i$  &$2+0.0i$ &$2+0.3i$ &$2+0.0i$ \\
4002 {\rm MeV}
&$0.3$ &$4+0.4i$   &$--$    &$--$        &$6+0.0i$  &$4+0.0i$ &$4+0.1i$ &$4+0.0i$   &$4+0.4i$&$4+0.0i$   \\
&$0.5$ &$10+0.7i$   &$--$    &$--$        &$12+0.0i$  &$9+0.0i$  &$9+0.2i$ &$9+0.1i$  &$10+0.7i$&$9+0.1i$  \\
\specialrule{0em}{1pt}{1pt}
$\Xi_c\Xi^*(2^+)$
&$0.2$ &$2+0.4i$   &$--$     &$--$      &$2+0.0i$  &$2+0.0i$  &$2+0.0i$  &$2+0.3i$ &$2+0.1i$ &$2+0.0i$ \\
4002 {\rm MeV}
&$0.3$ &$4+0.5i$   &$--$    &$--$       &$4+0.1i$  &$4+0.0i$ &$4+0.0i$  &$4+0.4i$  &$4+0.1i$&$4+0.0i$   \\
&$0.5$ &$10+1.0i$   &$--$    &$--$       &$9+0.2i$  &$9+0.0i$  &$9+0.1i$  &$10+0.8i$ &$11+0.3i$&$9+0.1i$  \\
\specialrule{0em}{1pt}{1pt}
$\Xi^*_c\Xi(1^+)$
&$1.5$ &$3+13.6i$   &$--$     &$--$      &$--$  &$1+0.0i$  &$1+0.0i$  &$7+14.3i$ &$1+0.1i$ &$1+0.1i$ \\
3963 {\rm MeV}
&$2.0$ &$7+17.0i$   &$--$    &$--$       &$--$  &$4+1.0i$ &$4+0.0i$  &$27+20.5i$  &$4+0.2i$&$4+0.2i$   \\
&$2.5$ &$15+15.0i$   &$--$    &$--$       &$--$  &$7+0.0i$  &$7+0.0i$  &$51+19.9i$ &$7+0.4i$&$7+0.4i$  \\
\specialrule{0em}{1pt}{1pt}
$\Xi^*_c\Xi(2^+)$
&$1.5$ &$0+14.0i$   &$--$     &$--$      &$--$  &$0+0.0i$  &$0+0.1i$ &$3+10.6i$  &$0+0.0i$ &$0+0.1i$ \\
3963 {\rm MeV}
&$2.0$ &$17+22.0i$   &$--$    &$--$        &$--$  &$4+0.5i$ &$1+0.3i$  &$15+18.1i$  &$1+0.1i$&$1+0.1i$   \\
&$2.5$ &$31+29.7i$   &$--$    &$--$       &$--$  &$6+0.0i$  &$3+0.6i$  &$31+22.0i$ &$2+0.2i$&$2+0.1i$  \\
\specialrule{0em}{1pt}{1pt}
$\Xi^{'}_c\Xi(0^+)$
&$0.8$ &$2+10.8i$   &$--$     &$--$      &$--$  &$--$  &$--$ &$1+0.7i$  &$1+6.9i$ &$1+0.0i$ \\
3896 {\rm MeV}
&$1.0$ &$11+18.6i$   &$--$    &$--$       &$--$  &$--$  &$--$  &$4+1.0i$  &$5+12.0i$&$2+0.0i$   \\
&$1.2$ &$18+16.2i$   &$--$    &$--$       &$--$  &$--$  &$--$ &$16+1.3i$  &$14+16.7i$&$4+0.0i$  \\
\specialrule{0em}{1pt}{1pt}
$\Xi^{'}_c\Xi(1^+)$
&$0.8$ &$2+10.1i$   &$--$     &$--$      &$--$  &$--$  &$--$  &$1+0.8i$ &$0+16.3i$ &$0+0.0i$ \\
3896 {\rm MeV}
&$1.0$ &$10+16.4i$   &$--$    &$--$       &$--$  &$--$  &$--$  &$3+1.2i$  &$4+10.9i$&$1+0.0i$   \\
&$1.2$ &$15+14.8i$   &$--$    &$--$       &$--$  &$--$  &$--$  &$5+1.6i$ &$9+15.2i$&$3+0.0i$  \\
\specialrule{0em}{1pt}{1pt}
\specialrule{0em}{1pt}{1pt}
$\Omega^*_c\Lambda(1^+)$
&$3.9$ &$1+0.5i$   &$--$     &$--$      &$--$  &$--$  &$--$ &$--$ &$1+0.0i$  &$1+0.6i$ \\
3882 {\rm MeV}
&$4.1$ &$5+1.1i$   &$--$    &$--$            &$--$  &$--$    &$--$  &$--$&$3+0.0i$&$6+1.4i$   \\
&$4.2$ &$8+1.4i$   &$--$    &$--$         &$--$  &$--$  &$--$ &$--$ &$5+0.0i$ &$9+1.7i$  \\
\specialrule{0em}{1pt}{1pt}
$\Omega_c\Lambda(1^+)$
&$4.1$ &$2+2.5i$   &$--$     &$--$      &$--$  &$--$  &$--$ &$--$ &$--$  &$2+2.5i$ \\
3810 {\rm MeV}
&$4.4$ &$3+3.5i$   &$--$    &$--$            &$--$  &$--$    &$--$  &$--$&$--$&$3+3.5i$   \\
&$4.6$ &$11+5.8i$   &$--$    &$--$         &$--$  &$--$  &$--$ &$--$ &$--$ &$11+5.8i$  \\
\specialrule{0em}{1pt}{1pt}
\bottomrule[1.5pt]
\end{tabular}
\end{table*}

Glancing over the coupled-channel results  of channels $\Xi^{(',*)}_c\Xi^{*}$,
$\Xi^{'}_c\Xi$ and $\Omega^{(*)}\Lambda$ in Table.~\ref{t0}, we can find  that
the real parts of most poles from the coupled-channel calculation are similar to
those from the single-channel calculations, and the small widths are acquired
from the couplings with the channels considered.  However, it has a great impact
on the $\Xi^{*}_c\Xi$ channel after including the full coupled-channel
interactions as suggested by  the variation in the mass and width.  Compared with
single-channel calculations, the masses change significantly, and the widths are
much larger.   Two-channel calculations are also performed, and the results are
presented in the fourth to eleventh columns. For the states near the
$\Xi^{(',*)}_c\Xi^{(*)}$  threshold with $(0,1,2,3)^+$, relatively obvious
two-channel couplings can be found in the $\Xi^{'}_c\Xi$ channel.  For the
states near the $\Xi^{'}_c\Xi^{*}$ threshold with  $(1,2)^+$, the main
two-channel couplings can be found in the $\Xi^{'}_c\Xi$ channel. For two states
near the $\Xi_{c}\Xi^{*}$ threshold with $(1,2)^+$, the widths from two-channel
couplings are both less than 1.0~MeV.  For the states near the $\Xi^{*}_c\Xi$
threshold with $(1,2)^+$, the main decay channel are $\Omega^{*}_c\Lambda$,
which leads to a width of about a dozen of MeVs and large increase of binding
energy.  Similarly, the states near the $\Xi^{'}_c\Xi$ threshold  with $(0,1)^+$
have considerable large couplings with the $\Omega_c\Lambda$ channel, which
leads to obvious increase of mass.  For the state near the  $\Omega^{*}_c\Lambda$
threshold with $1^+$, the $\Xi_c\Xi$ channel is the dominant channel to produce
their total widths.  Since the $\Omega_c\Lambda$ channel has the second lowest
threshold, it can only couple to the $\Xi_c\Xi$ channel so that the only
two-channel coupling width came from the $\Xi_c\Xi$ channel.

\renewcommand\tabcolsep{0.17cm}
\renewcommand{\arraystretch}{1.1}
\begin{table*}[hpbt!]   %hpbt!
\footnotesize
\caption{The masses and widths of isovector charmed-strange molecular states at different values of $\alpha$. Other notations are the same as Table~\ref{t0}.  \label{t1}}
%\caption{Coupled channel results of the baryon-baryon states with $I=1$\label{t1a}}
\begin{tabular}{crrrrrrrrrrrrrrrr}\bottomrule[1.5pt]
\specialrule{0em}{1pt}{1pt}
$I=1$&$\alpha_r$ &\multicolumn{1}{c}{$CC$}& \multicolumn{1}{c}{  $\Omega^*_c\Sigma^*$} &\multicolumn{1}{c}{  $\Sigma_c\Omega$}&\multicolumn{1}{c}{$\Xi^{'}_c\Xi^*$}& \multicolumn{1}{c}{$\Omega_c\Sigma^*$} & \multicolumn{1}{c}{$\Xi_c\Xi^*$} & \multicolumn{1}{c}{$\Xi^{*}_c\Xi$}& \multicolumn{1}{c}{  $\Omega^*_c\Sigma$} & \multicolumn{1}{c}{$\Xi^{'}_c\Xi$} & \multicolumn{1}{c}{$\Omega_c\Sigma$}& \multicolumn{1}{c}{$\Xi_c\Xi$} \\
\specialrule{0em}{1pt}{1pt}
\hline
\specialrule{0em}{1pt}{1pt}
$\Xi^*_c\Xi^*(0^+)$
&$0.6$ &$2+7.6i$&$1+3.8i$ &$1+0.0i$&$1+0.0i$&$1+0.1i$  &$1+0.0i$  &$1+0.0i$ &$1+0.0i$ &$1+0.1i$ &$1+0.1i$ &$1+0.1i$\\
4178 {\rm MeV}
&$0.8$ &$8+6.0i$   &$3+7.4i$ &$4+0.0i$&$4+0.0i$  &$4+0.1i$  &$4+1.5i$  &$4+0.0i$  &$4+.0i$ &$4+0.1i$&$4+0.2i$&$4+0.1i$   \\
&$1.0$ &$18+0.7i$   &$5+13.6i$  &$7+0.0i$&$7+0.0i$  &$8+0.2i$  &$8+2.7i$  &$7+0.0i$ &$7+0.1i$ &$7+0.2i$&$8+0.3i$&$7+0.2i$  \\
\specialrule{0em}{1pt}{1pt}
$\Xi^*_c\Xi^*(1^+)$
&$0.7$ &$1+5.5i$   &$0+4.3i$  &$1+0.0i$&$1+0.0i$  &$1+0.0i$  &$1+0.2i$ &$1+0.0i$ &$1+0.1i$ &$1+0.1i$&$1+0.1i$&$1+0.1i$ \\
4178 {\rm MeV}
&$0.9$ &$6+6.8i$   &$2+5.9i$  &$5+0.0i$ &$5+0.0i$ &$5+0.0i$   &$5+0.3i$  &$5+0.1i$ &$5+0.2i$&$5+0.1i$&$5+0.2i$ &$5+0.1i$   \\
&$1.1$ &$10+3.6i$  &$5+10.0i$ &$7+0.0i$&$7+0.0i$ &$7+0.0i$  &$7+0.4i$ &$7+0.1i$ &$7+0.2i$&$7+0.1i$ &$7+0.3i$ &$6+0.2i$ \\
\specialrule{0em}{1pt}{1pt}
$\Xi^*_c\Xi^*(2^+)$
&$1.0$ &$0+4.1i$  &$0+2.0i$ &$2+0.0i$&$2+0.1i$ &$2+0.7i$  &$2+0.5i$ &$2+0.1i$ &$2+0.3i$ &$2+0.0i$&$2+0.2i$&$2+0.1i$ \\
4178 {\rm MeV}
&$1.2$ &$1+5.8i$   &$1+3.0i$ &$5+0.0i$ &$5+0.2i$&$3+0.9i$  &$4+0.9i$  &$5+0.2i$ &$5+0.5i$&$5+0.1i$ &$5+0.3i$&$5+0.1i$   \\
&$1.4$ &$3+9.7i$   &$3+5.0i$&$7+0.0i$ &$7+0.3i$&$6+1.1i$  &$7+1.5i$ &$7+0.4i$ &$8+0.8i$&$7+0.1i$&$7+0.5i$&$7+0.2i$  \\
\specialrule{0em}{1pt}{1pt}
$\Omega^*_c\Sigma^*(0^+)$
&$3.7$ &$0+9.7i$   &$--$ &$--$ &$15+7.6i$ &$1+0.0i$  &$0+4.5i$  &$0+18.5i$  &$0+0.0i$   &$1+0.5i$&$1+0.0i$&$22+1.5i$   \\4150 {\rm MeV}
&$3.9$ &$7+11.6i$   &$--$ &$--$ &$21+8.6i$ &$2+0.0i$  &$6+5.2i$  &$0+27.1i$  &$0+0.0i$   &$2+0.5i$&$2+0.0i$ &$25+5.5i$ \\
&$4.1$ &$24+13.2i$   &$--$ &$--$ &$27+8.9i$ &$4+0.0i$  &$12+7.2i$  &$3+34.0i$  &$0+0.0i$ &$4+0.4i$&$4+0.0i$&$27+9.7i$   \\
&$4.7$ &$27+18.5i$   &$--$ &$--$ &$38+0.0i$ &$13+0.0i$  &$23+13.1i$  &$29+44.6i$  &$1+0.1i$ &$13+0.0i$&$13+0.0i$&$28+21.5i$   \\
\specialrule{0em}{1pt}{1pt}
$\Omega^*_c\Sigma^*(1^+)$
&$2.0$ &$0+13.1i$   &$--$ &$--$    &$8+9.4i$      &$0+0.0i$  &$0+44.9i$  &$0+3.2i$ &$21+24.2i$ &$0+0.4i$ &$0+0.0i$ &$0+0.1i$\\
4150 {\rm MeV}
&$2.2$ &$4+15.6i$   &$--$ &$--$   &$30+7.7i$       &$1+0.0i$  &$16+31.8i$  &$0+5.1i$  &$38+29.0i$ &$0+3.0i$&$1+0.0i$&$0+0.1i$   \\
&$2.4$ &$28+21.2i$   &$--$ &$--$   &$40+0.0i$       &$9+0.0i$  &$27+20.6i$  &$3+17.3i$  &$48+33.5i$&$4+6.2i$ &$9+0.0i$&$8+1.0i$ \\
\specialrule{0em}{1pt}{1pt}
$\Omega^*_c\Sigma^*(2^+)$
&$2.1$ &$2+4.6i$   &$--$&$--$&$0+0.2i$    &$1+0.0i$   &$0+0.0i$ &$0+6.0i$ &$27+32.7i$  &$0+3.5i$  &$1+0.0i$ &$0+0.6i$  \\4150 {\rm MeV}
&$2.5$ &$15+7.4i$   &$--$ &$--$ &$0+0.4i$  &$18+0.0i$  &$0+42.0i$&$0+21.5i$  &$49+59.1i$        &$0+15.1i$ &$18+0.0i$&$14+2.5i$  \\
&$2.9$ &$29+14.1i$   &$--$ &$--$ &$4+4.8i$  &$45+0.0i$  &$7+40.2i$ &$2+17.9i$ &$--$  &$6+30.3i$ &$44+0.0i$&$38+4.2i$  \\
\specialrule{0em}{1pt}{1pt}
$\Omega^*_c\Sigma^*(3^+)$
&$2.2$ &$5+16.8i$   &$--$ &$--$    &$0+0.0i$ &$3+0.0i$  &$0+0.4i$ &$--$ &$12+27.5i$ &$0+5.1i$ &$3+0.0i$ &$0+2.0i$\\4150 {\rm MeV}
&$2.4$ &$19+18.9i$   &$--$&$--$    &$0+0.0i$       &$8+0.0i$   &$0+0.5i$ &$--$  &$27+39.6i$ &$0+11.2i$&$8+0.0i$&$0+3.0i$   \\
&$2.6$ &$29+24.7i$   &$--$ &$--$   &$0+0.2i$       &$16+0.0i$  &$0+0.5i$ &$--$ &$45+57.1i$ &$0+20.9i$&$16+0.0i$ &$0+3.8i$ \\
&$2.8$ &$36+23.6i$   &$--$ &$--$   &$3+0.6i$      &$26+0.0i$  &$1+4.9i$ &$2+0.8i$ &$--$ &$3+32.1i$&$26+0.0i$ &$2+4.6i$ \\
\specialrule{0em}{1pt}{1pt}
$\Xi^{'}_c\Xi^*(1^+)$
&$0.4$ &$0+3.9i$   &$--$  &$--$   &$--$      &$0+3.9i$  &$1+0.2i$  &$1+0.0i$ &$1+0.1i$ &$1+0.0i$ &$1+0.1i$ &$1+0.0i$\\
4111 {\rm MeV}
&$0.8$ &$3+11.7i$   &$--$ &$--$   &$--$       &$1+15.4i$  &$8+0.3i$  &$7+0.1i$  &$8+0.2i$ &$7+0.1i$&$8+0.2i$&$7+0.1i$   \\
&$1.2$ &$6+19.2i$   &$--$  &$--$  &$--$       &$1+25.2i$  &$14+2.8i$  &$16+0.2i$ &$16+0.2i$ &$16+0.2i$&$15+1.4i$  &$16+0.3i$\\
\specialrule{0em}{1pt}{1pt}
$\Xi^{'}_c\Xi^*(2^+)$
&$0.4$ &$0+2.8i$   &$--$  &$--$   &$--$      &$0+5.8i$  &$1+0.2i$  &$1+0.0i$ &$1+0.0i$ &$1+0.0i$ &$1+0.1i$&$1+0.0i$ \\
4111 {\rm MeV}
&$0.8$ &$0+3.2i$   &$--$ &$--$   &$--$      &$0+2.4i$  &$7+1.0i$  &$7+0.1i$  &$7+0.1i$&$7+0.1i$&$7+0.4i$ &$7+0.0i$ \\
&$1.2$ &$3+7.8i$   &$--$ &$--$   &$--$      &$6+10.0i$  &$14+2.8i$  &$16+0.2i$  &$16+0.2i$&$16+0.2i$&$15+1.4i$ &$16+0.3
i$ \\
\specialrule{0em}{1pt}{1pt}
$\Omega_c\Sigma^*(1^+)$
&$2.4$ &$0+3.3i$   &$--$ &$--$ &$--$  &$--$  &$4+13.5i$  &$0+1.8i$  &$0+0.0i$      &$--$&$5+1.4i$&$0+2.6i$   \\4079 {\rm MeV}
&$2.6$ &$0+3.9i$   &$--$ &$--$ &$--$  &$--$  &$9+16.4i$  &$0+2.6i$  &$0+0.0i$     &$--$&$8+1.7i$&$0+3.3i$   \\
&$3.3$ &$3+7.5i$   &$--$ &$--$ &$--$  &$--$  &$32+19.8i$  &$1+5.7i$ &$1+0.0i$ &$--$&$28+2.7i$&$0+9.8i$   \\
&$3.5$ &$5+8.8i$   &$--$ &$--$ &$--$  &$--$  &$44+16.8i$  &$2+6.7i$ &$2+0.0i$ &$--$&$34+16.8i$  &$2+14.8i$\\
\specialrule{0em}{1pt}{1pt}
$\Omega_c\Sigma^*(2^+)$
&$1.4$ &$0+6.9i$   &$--$ &$--$   &$--$      &$--$  &$0+3.1i$  &$0+0.9i$ &$0+0.0i$ &$--$ &$0+13.9i$ &$0+0.7i$\\
4079 {\rm MeV}
&$1.6$ &$0+8.7i$   &$--$&$--$    &$--$      &$--$  &$5+7.5i$  &$2+3.0i$  &$2+0.0i$&$--$&$13+23.4i$ &$0+2.7i$  \\
&$1.8$ &$7+11.2i$   &$--$ &$--$   &$--$      &$--$  &$11+9.2i$  &$9+5.9i$  &$9+0.0i$&$--$&$31+33.0i$ &$0+5.9i$ \\
&$2.2$ &$16+17.8i$   &$--$ &$--$   &$--$      &$--$  &$18+10.9i$  &$31+14.1i$  &$30+0.0i$&$--$&$45+50.0i$ &$4+19.7i$ \\
\specialrule{0em}{1pt}{1pt}
$\Xi_c\Xi^*(1^+)$
&$0.4$ &$1+0.0i$   &$--$ &$--$    &$--$     &$--$  &$--$  &$1+0.0i$ &$1+0.0i$ &$1+0.0i$ &$1+0.0i$ &$1+0.0i$\\
4002 {\rm MeV}
&$0.6$ &$4+0.1i$   &$--$ &$--$   &$--$      &$--$  &$--$  &$3+0.0i$  &$3+0.0i$ &$3+0.0i$&$4+0.1i$&$3+0.0i$   \\
&$0.8$ &$7+0.2i$   &$--$ &$--$   &$--$      &$--$  &$--$  &$7+0.0i$  &$7+0.0i$&$7+0.0i$&$7+0.1i$ &$7+0.0i$ \\
\specialrule{0em}{1pt}{1pt}
$\Xi_c\Xi^*(2^+)$
&$0.4$ &$1+0.1i$   &$--$ &$--$    &$--$      &$--$  &$--$  &$1+0.0i$  &$1+0.0i$&$1+0.0i$ &$1+0.0i$&$1+0.0i$ \\
4002 {\rm MeV}
&$0.6$ &$4+0.2i$   &$--$ &$--$   &$--$       &$--$  &$--$   &$3+0.0i$   &$4+0.0i$&$3+0.1i$&$3+0.0i$ &$3+0.0i$  \\
&$0.8$ &$7+0.4i$   &$--$ &$--$   &$--$       &$--$  &$--$  &$7+0.0i$ &$7+0.1i$ &$7+0.1i$&$7+0.1i$  &$7+0.0i$\\
\specialrule{0em}{1pt}{1pt}
$\Xi^*_c\Xi(1^+)$
&$0.6$ &$2+0.2i$   &$--$  &$--$   &$--$      &$--$  &$--$  &$--$  &$1+0.0i$&$2+0.0i$ &$1+0.0i$&$1+0.0i$ \\
3963 {\rm MeV}
&$0.8$ &$6+0.1i$   &$--$ &$--$   &$--$       &$--$  &$--$  &$--$  &$3+0.0i$ &$5+0.0i$&$4+0.1i$&$3+0.0i$   \\
&$1.0$ &$10+0.3i$   &$--$ &$--$   &$--$       &$--$  &$--$  &$--$  &$7+0.0i$ &$7+0.1i$&$7+0.1i$  &$7+0.0i$\\
\specialrule{0em}{1pt}{1pt}
$\Xi^*_c\Xi(2^+)$
&$0.6$ &$2+0.2i$   &$--$  &$--$   &$--$      &$--$  &$--$  &$--$ &$1+0.0i$ &$1+0.0i$ &$1+0.0i$&$1+0.0i$ \\
3963 {\rm MeV}
&$0.8$ &$5+0.2i$   &$--$ &$--$   &$--$       &$--$  &$--$    &$--$   &$4+0.0i$&$4+0.1i$&$4+0.1i$ &$4+0.0i$  \\
&$1.0$ &$9+0.4i$   &$--$ &$--$   &$--$       &$--$  &$--$  &$--$ &$7+0.1i$ &$8+0.2i$&$7+0.1i$&$8+0.0i$  \\
\specialrule{0em}{1pt}{1pt}
$\Xi^{'}_c\Xi(0^+)$
&$0.4$ &$4+0.2i$   &$--$ &$--$    &$--$      &$--$  &$--$  &$--$ &$--$&$--$ &$4+0.2i$ &$3+0.0i$\\
3896 {\rm MeV}
&$0.6$ &$8+0.0i$   &$--$&$--$    &$--$       &$--$  &$--$    &$--$  &$--$&$--$&$8+0.0i$&$7+0.0i$   \\
&$0.8$ &$14+0.0i$   &$--$ &$--$   &$--$       &$--$  &$--$  &$--$ &$--$&$--$&$14+0.0i$ &$12+0.0i$ \\
\specialrule{0em}{1pt}{1pt}
$\Xi^{'}_c\Xi(1^+)$
&$0.4$ &$3+0.2i$   &$--$ &$--$    &$--$      &$--$  &$--$  &$--$ &$--$&$--$ &$3+0.2i$ &$3+0.0i$\\
3896 {\rm MeV}
&$0.6$ &$8+0.0i$   &$--$ &$--$   &$--$        &$--$  &$--$    &$--$ &$--$ &$--$&$8+0.6i$&$7+0.0i$   \\
&$0.8$ &$13+0.1i$   &$--$ &$--$   &$--$        &$--$  &$--$  &$--$ &$--$&$--$&$13+0.0i$ &$12+0.0i$ \\
\specialrule{0em}{1pt}{1pt}
\bottomrule[1.5pt]
\end{tabular}
\end{table*}

The coupled-channel results of isovector baryon-baryon interactions are
presented in Table ~\ref{t1}. For the isovector states near the
$\Xi^{*}_c\Xi^{*}$ threshold with $(0,1,2)^+$, large couplings can be found in
the $\Omega^{*}_c\Sigma^{*}$ channel and their binding energies also decrease a
little compared with the single-channel results after including  the two-channel
couplings.  Among the states near the $\Omega^{*}_c\Sigma^{*}$ threshold with
$(0,1,2,3)^+$, there exist some differences between different two-channel
couplings. After including the two-channel couplings between the channel
$\Omega^{*}_c\Sigma^{*}$ and the channels $\Xi^{'}_c\Xi^*$, $\Xi_c\Xi^*$ or
$\Xi_c\Xi^*$, the binding energy of the state with $0^+$ becomes obviously
larger than the single-channel value together with considerable widths. The
coupling to the $\Omega_c^{*}\Sigma$ channel leads to a decrease of the binding
energy. Other two-channel couplings affect a little on the single-channel
results in mass and lead to small widths. For the state with $1^+$, large
couplings can be found in the channels $\Xi^{'}_c\Xi^*$ and $\Omega^{*}_c\Sigma$
with large widths.  However, when it couples to channels $\Omega_c\Sigma^{*}$,
$\Xi^{*}_c\Xi$ or $\Xi_c\Xi$, the bound state appears at a large $\alpha$ value
of about 2.4. The state with $2^+$ strongly couples to channel
$\Omega^{*}_c\Sigma$, and the couplings with channels $\Xi^{'}_c\Xi^{*}$,
$\Xi_c\Xi^{*}$, $\Xi^{'}_c\Xi$ or $\Xi_c\Xi$ result in decreases of the binding
energy. When the $(2,3)^+$ states couple to the channel $\Omega^{*}_c\Sigma$ at
the parameters 2.9 and 2.8, respectively, the two "$--$" in table mean the binding
energies beyond our coupled-channel calculation range with binding energy less
than $50$~MeV. For the isovector states near the $\Xi^{'}_c\Xi^{*}$ threshold
with $(1,2)^+$, the coupling effects have no significant effect compared with
the single-channel results as suggested by the almost unchanged masses and very
small widths.  However, the coupling effects decrease the binding energy and
brings considerable widths when they couple to the $\Omega_c\Sigma^{*}$ channel.
Hence, the two-channel results with the channel $\Omega_c\Sigma^{*}$, to some
extent, affect the overall coupled-channel results a lot and give rise to the
noticeable reduction in binding energies.  For the states near
$\Omega_c\Sigma^{*}$ threshold with $(1,2)^+$, the channels $\Xi_c\Xi^{*}$ and
$\Omega_c\Sigma$ are dominant. In addition, the states with $(1,2)^+ $ are not
attractive enough to be produced within the range of parameter value considered
after coupling to the $\Xi^{'}_c\Xi$ channel.  No obvious strongly
coupled-channel effects can be found for the left states near the
$\Xi_c\Xi^{*}$, $\Xi^{*}_c\Xi$ and $\Xi^{'}_c\Xi$ thresholds, and the width from
the two-channel couplings are all less than 1~MeV.

\section{Summary and discussion}\label{Sec: summary}

In this work, we systematically study the charmed-strange baryon systems
composed of $csssqq$ quarks and their baryon-antibaryon partners, in a qBSE
approach. The potential kernels are constructed with the help of the effective
Lagrangians with SU(3), chiral and heavy quark symmetries.  The S-wave bound
states are searched for as the pole of the scattering amplitudes.  All S-wave
charmed-strange dibaryon interactions $\Xi^{(',*)}_{c}\Xi^{(*)}$,
$\Omega^{(*)}_c\Lambda$, $\Omega^{(*)}_c\Sigma^{(*)}$, $\Lambda_c\Omega$ and
$\Sigma^{(*)}_c\Omega$ and their baryon-antibaryon partners
$\Xi^{(',*)}_{c}\bar{\Xi}^{(*)}$, $\Omega^{(*)}_c\bar{\Lambda}$,
$\Omega^{(*)}_c\bar{\Sigma}^{(*)}$, $\Lambda_c\bar{\Omega}$ and
$\Sigma^{(*)}_c\bar{\Omega}$ are considered, which leads to 84 channels with
different spin parities.

The single-channel calculations suggest that 36 and 24 bound states can be
produced from the baryon-baryon and baryon-antibaryon interactions,
respectively. Most bound states from  baryon-antibaryon interactions are
produced at much larger values of parameter $\alpha$, which suggests that these
bound states are less possible to be found in future experiments than corresponding dibaryon states. Such results are consistent with our previous
results~\cite{Kong:2022rvd} that fewer states can be produced in the
charmed-antistrange interaction than charmed-strange interactions.

Furthermore, the coupling effects on the produced bound states in the
single-channel calculations are studied.  Since the states from the
baryon-antibaryon interactions are less possible to exist, we do not consider
these interactions in coupled-channel calculations. For the isoscalar
interactions, the coupled-channel calculations hardly change the conclusion from
the single-channel calculations, which means that the coupled-channel effects
are not very significant. However, for the isovector interactions, the
coupled-channel effects have obvious effects, which usually cause great
variations of binding energy together with considerable widths. Compared with
our previous coupled-channel calculations in
Refs.~\cite{Zhu:2020vto,He:2019rva}, the coupled-channel effect has obvious
large influence on both the real part and imaginary part of poles. It may be
related to the constituent hadrons considered in the current work. The systems
studied in the current work are composed of a light hadron and a charmed hadron.
Compared with the double-charmed or double-bottom systems, the systems
containing light hadrons are usually more unstable.

Generally speaking, the charmed-strange dibaryon systems with  $csssqq$ quarks are usually attractive enough to produce bound states, while their baryon-antibaryon partners are less or hardly attractive. Both theoretical and experimental studies are suggested to give more valuable information.

\vskip 10pt

\noindent {\bf Acknowledgement} This project is supported by the Postgraduate
Research and Practice Innovation Program of Jiangsu Province (Grants No.
KYCX22\_1541) and the National Natural Science Foundation of China (Grants No.
11675228).

  \end{document}